\documentclass[11pt,twoside]{article}

\usepackage{asp2004}
\usepackage{epsf}
\usepackage{psfig}
\usepackage{lscape}

\begin{document}
\title{Formation and Evolution of Young Massive Clusters}    
\author{Bruce G. Elmegreen}   
\affil{IBM Research Division, T.J. Watson Research Center, 1101 Kitchawan Road,
Yorktown Hts., NY 10598, USA, bge@watson.ibm.com}    

\begin{abstract}
Clusters are the dense inner regions of a wide-spread hierarchy of
young stellar structures. They often reveal a continuation of this
hierarchy inside of them, to smaller scales, when they are young,
but orbital mixing eventually erases these subparts and a only
smooth cluster or smooth unbound group remains.  The stellar
hierarchy follows a similar structure in the interstellar gas,
which is presumably scale-free because of supersonic motions in
the presence of turbulence and self-gravity.  The efficiency of
star formation increases automatically with density in a
hierarchical ISM, causing most dense stellar groups to be
initially bound for local conditions. In lower pressure
environments, the infant mortality rates should be higher. Also
following from hierarchical structure is the cluster mass
distribution function and perhaps also the cluster size
distribution function, although the predicted mass-size relation
is not observed. Cluster destruction is from a variety of causes.
The destruction time should depend on cluster mass, but the
various groups who have studied this dependence have gotten
significantly different results so far.
\end{abstract}

\section{Introduction: Cluster Basics}

\subsection{Hierarchical structure in young star fields}

We know what a cluster is when we see one, but there is a lot more
to clustering than meets the eye. Embedded clusters often contain
sub-clusters, for example, and clusters generally cluster together
themselves into double clusters or star complexes. Taken together,
there is a hierarchy of young stellar structures, with the objects
commonly called clusters representing only the inner and denser
parts of the hierarchy. Presumably the main difference between the
clusters and the rest of the hierarchy is that the clusters have
had sufficient time and gravitational self-attraction to get mixed
by stellar orbital motions. The rest of the hierarchy could
partially mix later, by cluster coalescence, for example, or it
could disburse through tidal forces.

The subclustering of clusters is evident when the clusters are
still young in terms of their dynamical time. Examples are NGC
2264, which has four subclusters with slightly different ages
($\pm1$ My; Dahm \& Simon 2005), rho Oph (Smith et al. 2005), and
Serpens (Testi et al. 2000).  Not all clusters will be born with
clear sub-clustering. If a cloud core has a strong radial density
gradient, then primordial gas subclustering can be erased or mixed
by tidal forces (for the same reason that clouds with strong
density gradients will not fragment much during free fall
collapse). However, cloud cores with strong density gradients are
already fairly old when measured in gas dynamical times; i.e.,
they have had at least a crossing time to respond to
self-gravitational forces.  In these cases, the hierarchy of the
gas is smoothed over somewhat by dynamical motions even before
star formation begins. Nevertheless, the cluster that results is
the mixed inner region of the hierarchy, with the mixing occurring
mostly in the gas phase rather than the stellar phase in this case
with a strong gradient.

On larger scales, young stars cluster together in unbound
structures, like OB associations. Efremov (1995) has studied such
large scale clustering for 30 years.  Most recently, Ivanov (2005)
mapped the star complexes in M33 and Bastian et al. (2005) studied
cluster complexes in M51. There is no characteristic length or
mass scale for these larger structures; the distribution functions
for their luminosity and size are power laws (Elmegreen et al.
2006, and references therein). An OB association has a
characteristic size of about 80 pc (Efremov 1995), but this is
only because of a selection effect: There is a general correlation
between the duration of star formation and the size of the region,
and OB associations are selected as concentrations of OB stars.
This limits their age to $\sim10$ My and thereby limits their size
to the observed 80 pc (Efremov \& Elmegreen 1998). Collections of
older stars, such as red supergiants or Cepheid variables, are
larger, 650 pc according to Efremov (1995). Collections of younger
stars, such as pre-main sequence stars, are smaller (in embedded
clusters).  The origin of the correlation is probably turbulence,
because the ratio of the size to the age, which is a velocity,
correlates with the size like the linewidth-size relation for
molecular clouds, having about the same slope and intercept
(Elmegreen 2000).

Clusters have been defined {\it historically} as small,
gravitationally bound, isolated collections of stars; their
density exceeds the local tidal density. By the time the objects
appear as ``clusters,'' the stars in them are mixed along with
most previous subclusterings, and the peripheral stars are
dispersed. Thus they tend to look isolated and unique. In fact
they are not born that way, they are born as part of a hierarchy
of young stellar structures without much distinction other than
their location in the high density parts.

Simulations of collapsing clouds by Bonnell, Bate \& Vine (2003)
show hierarchical star formation with clusters in several dense
cores.  These clusters mix together over time making larger
clusters. The observations do not yet clearly distinguish between
clouds like these that are collapsing freely and clouds that have
some level of radial stability resulting from magnetic forces or
isotropic turbulent motions.  Quasi-stability seems favored by
observations of cloud core self-absorption where the absorption
line redshift is only $\sim20$\% of the linewidth (Myers et al.
2000). Such small redshifts suggest that cloud cores are
contracting relatively slowly, on a time scale of several
radial-crossing times. However, recent observations of global
infall in NGC 1333 (Walsh, Bourke, \& Myers 2006) suggest
dynamical processes on shorter times, similar to those envisioned
by Bonnell, Bate \& Vine (2003). In either case, the stars should
form in a hierarchical way (Sect. \ref{sect:hier}).

\subsection{Hierarchical structure in the gas}
\label{sect:hier2}

Interstellar gas has a similar hierarchy of structures. Power
spectra of gas emission show power laws, indicating no
characteristic scale (Dickey et al. 2001).  The power spectra of
optical light in a galaxy is about the same as the power spectrum
of interstellar gas (Elmegreen et al. 2003). These results suggest
that the stars follow the gas when they form, although extinction
may contribute to the optical power spectrum too. In terms of
cloud-like objects, the highest level in the hierarchy, on the
largest scale, consists of giant atomic clouds that contain
$\sim10^7$ M$_\odot$ of gas. These are evident in the Milky Way
(McGee \& Milton 1964; Elmegreen \& Elmegreen 1987) and in local
galaxies where HI maps have the resolution needed to see them
(e.g., Boulanger \& Viallefond 1992).  Molecular clouds are the
next denser level in the hierarchy of structures, and they often
collect together inside the atomic clouds if the ambient pressure
is large enough to allow molecules to form (Elmegreen 1993; Blitz
\& Rosolowsky 2006). Maps in Grabelski et al. (1987) show this GMC
collection clearly for the Milky Way. Inside giant molecular
clouds are cloud cores, and so on down to the smallest scales that
can be observed.

It is worth making the cautionary note that this hierarchical
structure for both gas and stars does not mean every small object
is inside a larger one. Small objects can be to the side of large
objects, with no obvious larger scale structures around them. Most
objects do contain substructures, however, down to or below the
scale where thermal motions dominate (or random stellar motions
dominate in the case of stellar clustering). An example of this
type of structure is illustrated in Figure 1, below.

\subsection{Star Formation Efficiency}

Individual stars form at the bottom of the hierarchy, where the
gas is in the form of dense clumps. As we consider lower and lower
levels in the hierarchy, i.e., as the average gas density gets
higher, the mass fraction in the form of the dense star-forming
clumps goes up.  On the scale of the clumps themselves, the mass
fraction is unity. On the scale of the GMC, the mass fraction is
small because there is a large volume of low-density gas between
the cores and even more between the clumps. This is the nature of
the hierarchy, which is fractal. Stars form inside each clump at
some average efficiency, which may be about constant, perhaps
one-third or one-half (e.g., Matzner \& McKee 2000). Thus the
overall efficiency of star formation, which is the mass fraction
of a cloud that turns into stars, increases towards higher gas
density inside the cloud. This is why the efficiency in a cloud
core may be 20\% or 30\%, but the efficiency for a whole OB
association is only 5\%.  The OB association is the result of star
formation inside the densest clumps of a GMC, but there is a lot
of low-density intercore and interclump gas in a GMC that does not
form stars.

We see now another aspect of cluster formation: in the dense
regions of the hierarchy where clusters form (GMC cores), the mass
fraction in the form of individual star-forming clumps is
automatically high, so the efficiency of star formation is
automatically high. This means that the stars have a high probably
of ending up gravitationally bound together. High average density
and boundedness go together; one follows from the other in a
hierarchical interstellar medium.  The high efficiency needed for
bound cluster formation is not the result of special circumstances
related to the stars themselves, such as feedback processes, but
only the result of hierarchical gas structure. At the density that
is high enough to form a compact region of stars, one that stands
out to the eye as a star cluster, the efficiency of star formation
is automatically high.

\subsection{Theory of Cluster Formation}

Putting these concepts together, we can get most of the way toward
explaining the origin of star clusters: they are the inner mixed
regions of the young-star hierarchy, where the dynamical time is
short and the age is comparable to or longer than this dynamical
time (to ensure mixing). Because most star formation occurs in
only a few dynamical times, a high fraction of the clumps will
have formed stars by the time mixing has occurred, and because the
clumps represent a high fraction of the local gas mass in a cloud
core, the efficiency of star formation is generally high there.
Then the mixed stellar regions remain gravitationally bound after
the gas leaves. Thus bound clusters are an inevitable result of
star formation in hierarchically structured gas for ISM properties
like those in the Solar neighborhood.

\subsection{Bound versus unbound clusters}

This initial binding does not mean that young clusters stay bound
for long. Ninety percent lose a high fraction of their stars
within the first 10 My (Lada \& Lada 2003; Fall, Chandar \&
Whitmore 2005), leaving only small bound cores (e.g., Kroupa,
Aarseth \& Hurley 2001) or perhaps no bound cores at all.  Some
giant star forming regions, such as NGC 604 in M33, appear to have
no clusters or cluster remnants, as if all star formation were
isolated or initially unclustered. Ma\'iz-Apell\'aniz (2001)
studied several other super-OB associations of this type. Such
diversity in initial clustering follows from the hierarchical
model if the average cloud density varies. Stars presumably form
in similar cores everywhere, with pre-collapse densities of
$10^5-10^8$ cm$^{-3}$, but where the average density is low
because of a low pressure ISM, for example, then the mass fraction
of star-forming clumps can be low even in the densest regions of
molecular clouds (which are not very dense in this case).  Stars
will still formed clustered, and the efficiency will still be
highest in the core regions, but this peak efficiency may be only
$\sim10$\% rather than $\sim50$\%. In that case there is not
enough gravitational binding from stars alone to make a bound
cluster when the gas leaves. Thus the hierarchical model predicts
that the fraction of stars forming in initially bound clusters
should depend on the local ISM pressure or density. This
prediction is consistent with the observation by Larsen \&
Richtler (2000), who found that the fraction of uv light from
stars in clusters increases with star formation rate, considering
that star formation rate increases with gas column density and
therefore ISM pressure.

\subsection{Origins of Hierarchical Structure}
\label{sect:hier}

The origin of hierarchical structure in interstellar gas is
probably a combination of gravitational fragmentation and
turbulence compression. Both produce scale-free density structures
in regions where the total energy density is much larger than the
thermal pressure. This inequality holds for most of the diffuse
interstellar medium and much of the self-gravitating ISM because
collisional cooling rates are high enough, and background heating
rates are low enough, that the temperatures and thermal pressures
are usually very low. Motions from various kinematic energy
sources are then supersonic, and the mixture of these motions,
particularly with the velocity-size correlation that results,
produces correlated density structures. In the high density parts
of these structures, where self-gravity exceeds the background
tidal force and the high column density provides some shielding
from both radiative and kinematic energies, the gas has time and
freedom to collapse into stars. The correlated density ensures
that most of these collapses will be clustered together, in
patches of various size, and it also provides the self-gravity
that forces orbital mixing and self-binding to make a homogeneous
embedded young cluster.

As mentioned in Section \ref{sect:hier2}, not every cloud is
surrounded by a larger cloud in a Russian doll pattern. Some small
clouds are to the side of larger clouds and others may be between
large clouds, with no obvious connection to one or the other. When
clouds get pushed around by directed stellar pressures, the
structures that form by gravity and turbulence get modified and
cloud pieces can scatter anywhere. In this sense, it is not
uncommon for stars to form in isolation or in small unbound groups
that are far from clusters and associations.  It appears for our
Galaxy that the fraction of stars forming in this isolated fashion
is small, perhaps 10\% or less (Carpenter 2000). Recent Spitzer
Space Telescope observations of the Perseus cloud suggest that
$\sim60$\% of Class I protostars are outside the massive cores;
many of these could still be in low-mass subclusters (Jorgensen et
al. 2006).

\subsection{Hierarchical Structure in the Galaxy NGC 628}

Figure 1 shows a fractal Brownian motion model of a face-on galaxy
compared to observations of the cluster size function in the
galaxy NGC 628. On the left, a grayscale image of a model galaxy
is shown, with four gray levels corresponding to different ratios
of the density to the peak density (this figure is explained more
completely and shown in color in Elmegreen et al. 2006). The
darkest regions around the edges have the lowest projected
densities, between 0.3 and 0.4 of the peak, the light gray regions
inside of these have projected densities between 0.4 and 0.5 of
the peak, whitish regions are between 0.5 and 0.6, and the dark
gray regions inside the white regions have the highest densities,
greater than 0.6 times the peak. The model is initialized in a 3D
cube where the density probability distribution function is
log-normal and the 3D power spectrum of density is approximately a
power law with a slope of $-11/3$. This power spectrum is
comparable to that for velocity and passive scalar density in
Kolmogorov incompressible turbulence (it is also the best fit to
the data compared to other power spectra that vary by $\pm1$ in
slope). The cube is then multiplied by a Gaussian on the line of
sight in order to simulate a projected galaxy disk.

The panel on the right of Figure 1 shows cumulative size
distribution functions for connected regions, or ``clouds'' that
were found objectively, for the same 4 levels of density threshold
relative to the peak. The lowest curve (i.e., having the smallest
count) is for the highest density threshold, greater than 0.6
times the peak.  The crosses are B-band observations of the size
distribution function for star-forming regions in an HST image of
NGC 628 (with arbitrary shifts in each axis to fit the model). The
physical scale ranges from 5 pc to 155 pc. The size distribution
is a power law with a slope of $-1.5$ on this cumulative plot. The
models match the observations. This cumulative size distribution
corresponds to the differential function $n(R)dR\propto
R^{-2.5}dR$.

The size distribution for loose stellar groupings in NGC 628 is
approximately a power law from 2.5 pc to 150 pc (Elmegreen et al.
2006).  This size range is larger than that for bound clusters, so
it is interesting to ask whether the size distribution for loose
groupings is a continuation of the size distribution for dense
clusters. There is preliminary evidence for this from the cluster
size distribution in Bastian et al. (2005), who studied M51.  They
derived a size function $n(R)dR\propto R^{-2.2}dR$, which is
similar to that on larger scales. The analogy between unbound
groupings and dense clusters is not straightforward, however,
because the unbound groupings have a mass that increases with size
approximately as $M\propto R^{1.5}$ (Elmegreen et al. 2006).  This
mass-size relation, along with the size distribution function, is
consistent with a hierarchical star distribution with projected
fractal dimension $D=1.5$.  That is, for a fractal, we expect
$M\propto R^D$ and $n(R)dR\propto R^{-D-1}$ (Mandelbrot 1983).
The problem with this is that the mass depends very little on
radius in the dense clusters studied by Bastian et al. (2005). The
lack of a correlation between mass and radius is also present in
the data in Testi, Palla \& Natta (1999) and Larsen (2004).

\begin{figure}
\plotone{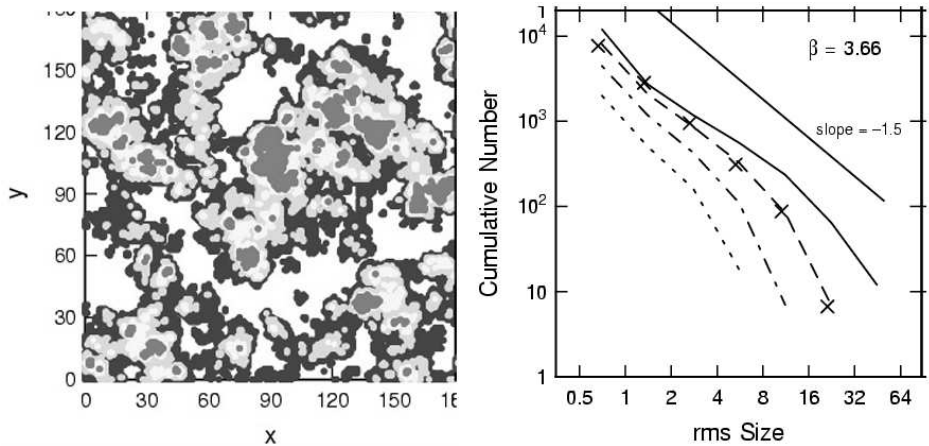} \caption{Fractal Brownian motion model
(left) of a projected galaxy and cumulative size distribution
functions (right) for clouds. The crosses are B-band data from HST
images of the galaxy NGC 628, corresponding to objects with
diameters ranging between 5 and 155 pc. The cumulative size
distribution is a power law with a slope of $-1.5$. The best model
has a density power spectrum comparable to the power spectrum of
Kolmogorov turbulence. (From Elmegreen et al. 2006.)}\end{figure}

\section{The Cluster Mass Function}

There are only a few observations of the mass distribution
functions of clusters.  There are many observations of luminosity
distribution functions, but mass distribution functions are more
difficult to obtain because they require the additional knowledge
of cluster ages. The complete mass function also requires the
conversion between luminosity and mass as a function of age. Here
we define the negative slope of the cluster mass function to be
$\beta$, so $M^{-\beta}dM$ is proportional to the number of
clusters with masses between $M$ and $M+dM$. A histogram of the
mass function plotted in equal intervals of $\log M$, which is the
most common way to display the mass function, would then have a
negative slope of $\beta-1$.

The most complete cluster sample for a nearby galaxy is in the
Large Magellanic Clouds. Elson \& Fall (1985) measured the cluster
luminosity function in the LMC and got $L^{-1.5}dL$ for a mixture
of ages. Elmegreen \& Efremov (1996) determined the luminosity
functions of LMC clusters in four age intervals and found them all
to be about $L^{-2}dL$. As the age was constant for each function,
mass is proportional to luminosity. This means the mass
distribution function is also a power law with the same slope,
$M^{-2}dM$, giving $\beta=2$.  A similar study of LMC clusters by
Hunter et al. (2003) found $\beta\sim2$, while de Grijs \& Anders
(2006) obtained $\beta=1.85\pm0.05$ using the same LMC data as in
Hunter et al. but different age calibrations.

Cluster luminosity functions in the Milky Way have been determined
for a long time, but not mass functions. One of the first modern
determinations was by Battinelli et al. (1994), who derived a
mass-luminosity relation for nearby Milky Way clusters and used
this to get a mass function slope of $\beta=2.13\pm0.15$ for the
high mass end of the cluster mass function. Battinelli et al. also
obtained $\beta=2.04\pm0.11$ for clusters in Lynga's catalogue.
For the Antenna galaxy, mass function slopes were found by Zhang
\& Fall (1999) to be $\beta=1.95\pm0.03$ for young clusters and
$\beta=2.00\pm0.08$ for old clusters. For M51, cluster mass
functions with $\beta=2$ and various upper mass cutoffs were
fitted to the observed luminosity functions in 3 HST passbands.
Similarly in NGC 3310, $\beta=2.04\pm0.23$ and in NGC 6745,
$\beta=1.96\pm0.15$ (de Grijs et al. 2003).  These observations
all imply that the cluster mass function is a power law with a
slope $\beta$ within several tenths of the value $2.0$.

The same slope $\beta=2$ is implied by galaxy-wide IMFs. Summed
IMFs from clusters can produce a global IMF that is the same as
the individual cluster IMF only if the cluster mass function slope
is $\beta=2$ or shallower. Observations suggest that galaxy-wide
IMFs are indeed the Salpeter IMF, not much steeper, based on
metallicity, color, H$\alpha$ equivalent width, and
color-magnitude diagrams (Elmegreen 2006a). This is the case for
the Milky Way bulge, dwarf galaxies, large galaxies, galaxy
clusters, and the whole Universe. Because cluster IMF slopes
average the Salpeter value too (Scalo 1998), the summed IMF equals
the cluster IMF and the cluster mass function has to be close to
$M^{-2}$. Even a slightly steeper value of $\beta=2.3$ makes the
summed IMF from clusters significantly steeper than the observed
IMF of composite populations (Kroupa \& Weidner 2003; Weidner \&
Kroupa 2005).

The cluster mass function is consistent with the theory of cluster
formation discussed in the previous section.  A hierarchical gas
distribution, forming clusters at arbitrary levels in the
hierarchy with a constant efficiency of star formation, has a mass
function slope of exactly $\beta=2$ (Fleck 1996). This can be seen
from a simple tree model where there is one trunk with a mass of
N=16, two branches off the trunk with masses of 8 each, two
branches off each branch with masses of 4 each, two more branches
off each of the previous with a mass of 2 each, and 16 total
branches at the top of the tree with a mass of 1 each. The product
of the number of branches times the mass of each branch is 16 for
all levels in the hierarchy. There are $1+2+4+8=15=N-1$ possible
branching points, each of which may be viewed as a possible
cluster with all of the smaller branches inside of that cluster.
If these 15 branching points are randomly sampled, then the
probability that the chosen point is the trunk, giving a
``cluster'' mass of 16, is $1/15$. The probability the chosen mass
is 8 is $2/15$, the probability it is 4 is $4/15$, and so on.
Considering the log of the mass in base 2, the probability that
the log is 4, so the mass is $2^4=16$, is $1/15$. The probability
the log is 3 is $2/15$, etc.. In general, the probability that the
log of the mass is $A$ is $2^{4-A}/15$, or, more generally,
$2^{N-A}/\left(2^N-1\right)$. This probability is proportional to
the count of branching points, and for the log mass expressions,
the counting is for equal intervals of log mass ($dA=d\log_2 M$).
Thus we have a probability function
\begin{equation}P\left(M=2^A\right)d\log_2 M =
2^{N-A}/\left(2^N-1\right)d\log_2 M.\end{equation} Converting $A$
on the right-hand side back to mass, we get
\begin{equation}P\left(M\right)d\log_2 M
={{2^N}\over{2^N-1}}M^{-1}d\log_2 M.\end{equation} This equation
shows that the mass distribution function of tree branches is
proportional to $M^{-1}$ for equal log intervals of mass, in which
case $\beta-1=1$ so $\beta=2$.  S\'anchez, Alfaro, \& P\'erez
(2006) show that the slope gets slightly shallower if blending
effects are considered.

A similar result has been obtained for a smooth gas density
distribution made by the fractal Brownian motion technique
(St\"utzke et al. 1998; Elmegreen 2002a; Elmegreen et al. 2006).
The mass function of three-dimensional clumps depends on the
contour level used to define the clump and is steeper for denser
levels. It is also steeper for steeper intrinsic power spectra
(St\"utzke et al 1998). It varies from $\beta\sim1.5$ for low
density to $\beta\sim2.3$ for high density when the fractal has a
power spectrum with a power law slope equal to that of a passive
scalar in incompressible Kolmogorov turbulence, and when the
density has a log-normal probability distribution function. This
variation in mass function slope with density is consistent with
the observation that the mass function for giant molecular clouds
is a shallow $\beta=1.5$ to 1.8, and the mass function for
clusters is steeper, $\beta=1.8$ to 2.1.  The difference is
presumably because whole GMCs sample a lower density in the ISM
than the star clusters they produce in their cores.

\section{Massive Star Formation in Clusters and in the Field}

An important question is whether massive stars can form in
isolation, either in the remote field or on the peripheral regions
of clusters.  If massive stars need the cluster environment to
accrete dense gas in a certain way, or to coalesce with other
protostars, then there should not be many forming in low density
regions (Testi, Palla, \& Natta 1999). Of course, it is understood
that all stars form in dense clumps, so the local environment is
never low density, but the question is whether each massive star
has a full complement of lower mass stars in the immediate
neighborhood, filling out the IMF toward lower mass. An isolated
dense clump could, in principle, form an isolated massive star
without the thousands of other stars expected from the usual IMF.

The discussion in the previous section suggested that each
logarithmic interval of cluster mass produces the same total
stellar IMF, regardless of the cluster mass itself. This is how
the summed IMF can equal the individual cluster IMF: low mass
clusters do not tip the balance toward exclusively low mass stars,
for example (Elmegreen 2006a). The implication is that stars of
any mass can form in clusters of any mass (provided the cluster
mass exceeds the stellar mass). One condition for this to be true
is that there is a universal IMF. If two or more star formation
processes operate differently in different regions and if one
tends to produce more high mass stars, then there are effectively
two or more IMFs that have various contributions to the total
depending on the environment (Elmegreen 2004). If the important
environmental variable is density, for example, then low density
clusters can have a slightly different IMF than high density
clusters. With such dichotomy, it would no longer be true that
stars of any mass form with equal probability in clusters of any
mass (or density).

A test of these two possibilities is the ``100-Taurus'' test. For
a universal IMF and $\beta\sim2$, the summed IMF from 100 separate
regions like the low mass Taurus clouds should have a slope that
is only $\sim0.1$ steeper than the IMF from one large region like
Orion, which contains 100 times the mass of Taurus. Observations
of IMFs in many low mass or low density regions like Taurus could
settle this issue directly. If a given number of stars in
Taurus-like regions have an IMF that is significantly steeper than
the IMF from the same number of stars in Orion, or in the Orion
Trapezium cluster, then there would seem to be at least two
distinct IMFs, and massive stars would be favoring the higher mass
and denser clusters.

At the moment, stars of any mass appear to form in clusters of any
mass. One can then think of a cluster as ``randomly'' sampling a
universal IMF.  For purely random sampling, there is a very small
chance that a massive star will form in a low-mass cluster, but
most massive stars form in high mass clusters because most stars
of all types form in high mass clusters. For the power-law part of
the IMF, the maximum likely star mass out of N clusters of mass M
equals the maximum likely star mass in one cluster of mass NM
(Elmegreen 2006a). These statements are consistent with the
observations by Oey, King \& Parker (2004), who found the
distribution of O-star counts per cluster in the LMC to be a power
law with $\beta\sim2$ down to a single O star. It is also
consistent with a stronger statement by de Wit et al. (2005), who
found that the total star mass in O-star containing clusters in
the solar neighborhood is a power law with $\beta=1.7$ down to a
single O star (not just an O-star cluster containing other stars).
This power law is similar to that for whole clusters, suggesting
that sometimes a single O star can form in place of a cluster that
has many smaller stars but the same total mass. The de Wit et al.
observation is consistent with 4\% of O-type stars forming in
isolation or in peripheral regions of clusters.

There are actually many examples where O-type stars form along the
periphery of massive dense clusters. Sequential triggering has
this effect, and in the 30 Dor region of the LMC it is
particularly clear (Walborn et al. 1999; see plot of massive stars
in Elmegreen 2006b).

\section{Cluster Disruption}

There are several reasons why clusters eventually come apart. Gas
dispersal in the first 1 to 3 My leads to decreased gravitational
binding for the initial stellar speed, and this causes some stars
to escape directly, in a crossing time or less (Lada, Margulis, \&
Dearborn 1984).  Many OB associations and loose stellar groups
could have been collections of embedded clusters several million
years earlier. Kroupa, Aarseth \& Hurley (2001) studied cluster
expansion after rapid gas loss with a star formation efficiency of
30\%. They found that as the cluster expands to a new equilibrium
radius, some stars are lost quickly and some remain in a bound
core. They suggested that a cluster like the Trapezium cluster in
Orion can turn into an open cluster like the Pleiades after an
amount of time has passed that is comparable to the age of the
Pleiades.

Mass is also lost from a cluster in the form of stellar winds and
supernovae during stellar evolution. It takes much longer for a
significant mass to be lost by stellar evolution than by cloud
core disruption. After a few tens of millions of years, the
decreasing cluster mass produces an overall expansion (Terlevich
1987). Eventually, the cluster density gets so low that
self-gravitational forces become comparable to or less than the
background tidal force. Then the cluster disperses.  This is the
third mechanism of cluster destruction. Tidal interactions with
dense clouds, spiral arms, the bulge, and the galactic disk all
give a lower limit to cluster density for survival. The outer
parts of the cluster are shed into a tidal tail at the radius of
this density threshold. As the cluster density decreases, the
tidal radius decreases too, so the bound part of the cluster
shrinks while the stars in the outer parts expand. Detailed models
of cluster disruption including evolution and tidal effects are in
Baumgardt \& Makino (2003).

Other talks at this conference consider cluster destruction in
greater detail.  In particular, the destruction of clusters by
giant molecular clouds has recently been shown to be important by
Gieles et al. (2006).  For molecular clouds, the important
quantity is the volume filling factor of molecular material with a
density comparable to or greater than the cluster density. When a
cluster enters this volume, it becomes tidally unbound for a time.
Movement near these dense regions can energize the stellar orbits,
leading to eventual destruction.  More distant encounters with
molecular clouds are less important than those with impact
parameters comparable to the cloud radius.  Intermediate mass
clusters may be disrupted by only a few GMC encounters, while low
and high mass clusters require many encounters.  The destruction
time for a cluster of mass $M$ through multiple encounters was
found to be
\begin{equation}t_{dis}=2\left({{\Sigma_{cloud}\rho_{cloud}}\over
{5.2\;{\rm M}_\odot^2\;{\rm
pc}^{-5}}}\right)^{-1}\left({{M}\over{10^4\;{\bf
M}_\odot}}\right)^{\gamma} \; {\rm Gyr}\end{equation} where
$\gamma=1-3\lambda$ and $\lambda$ is the power in the cluster
mass-radius relation, $R_{cluster}\propto M^\lambda$. The typical
mass column density of a molecular cloud is $\Sigma_{cloud}$ and
the ISM-averaged density of GMCs is $\rho_{cloud}$. Gieles et al.
(2006) suggest $\lambda\sim0.13$ and $\gamma\sim0.61$. Thus
$t_{dis}\propto M^{0.61}$.

A similar mass dependence for the destruction time of a cluster
has been found in other studies.  Boutloukos \& Lamers (2003)
found the mass-dependent disruption time in four galaxies to be
$t_{dis}\sim M^{0.6}$. Gieles et al. (2004) showed that models by
Baumgardt \& Makino (2003) were consistent with $t_{dis}\sim
M^{0.64}$. De la Fuente Marcos \& de la Fuente Marcos (2004) used
dynamical models to suggest $t_{dis}\sim M^{0.68}$. Most recently,
Lamers et al. (2005) compared $M(t)$ from numerical experiments
using this power law form for $t_{dis}=M^\gamma$ and the simple
model $dM/dt=- M/t_{dis}$. The model and analytical results were
in excellent agreement for $\gamma=0.62$.

If $t_{dis}$ depends on cluster mass, then the mass function of
clusters should get shallower at low mass over time because the
lowest mass clusters get destroyed soonest. After a while, the
slope of the cluster mass function should approach
$\beta=1-\gamma=0.38$ at low mass (Lamers et al. 2005).  Recall
that $\beta$ is defined to be the negative value of the slope on a
plot of the cluster mass spectrum in linear mass intervals, and
$\beta-1$ is the slope on a spectrum in log mass intervals. A
value of $\beta=0.38$ implies that a histogram of cluster number
in equal intervals of log mass should increase with $\log M$ at
low $M$, reach a peak, and then decrease with $\log M$ as
$M^{-\left(\beta-1\right)}$ for sufficiently large mass where
destruction has not been significant yet.  Most cluster mass
functions plotted with $\log M$ intervals are indeed peaked like
this, with a rising part at low $\log M$, but this is always
attributed to magnitude limits in the observations. That is, the
turnover toward low mass is the result of missing clusters that
are present but too faint to discern. For example, de Grijs \&
Anders (2006) plotted histograms of cluster counts in $\log M$
intervals for the LMC and found continuous power law functions
down to and below $10^4$ M$_\odot$ for ages up to $10^{9.75}$
years (this was also the case in Elmegreen \& Efremov [1996] and
Hunter, et al. [2003]).  There is no evidence for a mass-dependent
destruction in these data.

In contrast, Fall, Chandar \& Whitmore (2005) suggest that the
destruction rate is independent of cluster mass, and that the mass
function stays constant over time, which is consistent with the
observation by de Grijs \& Anders (2006) and also with
observations of the Antennae galaxy by Fall et al. Fall et al.
find that the number of clusters in intervals of equal age
decreases inversely with the age, independent of cluster mass.
This decrease is somewhat continuous over time all the way from
$10^6$ years to $10^9$ years. This is a surprising result because
there is not even a feature in this trend where the mechanism of
cluster disruption is expected to change from gas expulsion to
stellar evolution.

In the Fall et al. model, we can write the number of clusters more
massive than $M_0$ as $n_{M>M0}(t)\propto t^{-1}$. If the
observations also suggest that the form of the mass function does
not change over time and is $n(M)=n_0 M^{-2}$ for linear intervals
of $M$, then
\begin{equation}\int_{M0}^{\infty}n_0M^{-2}dM=n_0M_0^{-1}\propto
t^{-1}\end{equation} for constant lower detection limit $M_0$.
Thus $n_0\propto t^{-1}$, the maximum cluster mass is $\propto
t^{-1}$, and each cluster has its mass decrease as $t^{-1}$. We
also obtain that $t_{dis}\propto M^{-1}$ from the relations $dM/dt
\equiv-M/t_{dis}\propto  t^{-2}\propto M^2$, or $d\log M/d \log t
= -1$.

Why is there a difference between the Fall et al. (2006) model and
the Lamers et al. (2005) model? The key observation to distinguish
between these two models is the slope of the cluster mass function
versus time.

\section{Conclusions}

The ISM and star birth positions appear scale-free from the scale
of the Jeans length in the ambient ISM (on a kpc scale) to below
the star formation scale. This scale-free distribution is not
likely to be continuous in any one region, but still present on
average, and present in a piecewise sense.  Stars form in dense
clumps whose local mass fraction compared to the surrounding
molecular material is high and whose mixing time is relatively
short. The mixed young stars are the ``cluster.'' Gas removal
stops star formation in the dense core but not everywhere. It
often continues on the periphery of the cluster and in the
remaining molecular cloud as a result of sequential triggering.
This basic model explains the cluster mass spectrum, the galactic
star formation rate and star formation timescale (not discussed
here; see Elmegreen 2002b), and the observed stellar grouping
structures.  It does not give the independence between cluster
mass and size, however, which presumably involves additional
physical processes.

Cluster mass loss is dominated by gas disruption and stellar
evolution at first, by tidal shredding from giant molecular clouds
and spiral density waves after a while, and by thermal disruption
on the longest time scale. The disruption time varies with cluster
mass, but the exact scaling relationship is subject to debate.
Observations suggest that the power-law form of the cluster mass
distribution function is preserved for at least $10^9$ years in
some galaxy disks, in which case the disruption time is either
longer than this or it does not scale positively with mass in a
noticeable way.

\end{document}